# CA20108 COST Action: A Methodology for Developing FAIR Micrometeorological Networks

Branislava Lalic[1], Josef Eitzinger[2], Stevan Savić[3,4], Ana Firanj Sremac[1], Michael Scriney[5], Mark Roantree[6]

[1] University of Novi Sad, Faculty of Agriculture, Trg Dositeja Obradovića 8, 21000 Novi Sad, Serbia

[2] BOKU University, Department of Ecosystem Management, Climate and Biodiversity, Institute of Meteorology and Climatology, Gregor-Mendel Str. 33, A-1180 Vienna, Austria

[3] University of Novi Sad, Faculty of Sciences, Novi Sad Urban Climate Lab; Trg Dositeja Obradovića 3, 21000 Novi Sad, Serbia

[4] University of Banja Luka, Faculty of Natural Sciences and Mathematics; Dr. Mladena Stojanovića 2, 78000 Banja Luka, Bosnia & Herzegovina

[5] School of Computing, Dublin City University, Glasnevin, Dublin 9, Ireland

[6] Insight Centre for Data Analytics, Dublin City University, Glasnevin, Dublin 9, Ireland

**Abstract**

This article reports the outcomes of the FAIRNESS COST Action (CA20108), a coordinated European initiative aimed at advancing micrometeorological data toward compliance with the FAIR (Findable, Accessible, Interoperable, Reusable) principles. The article presents three core achievements: (i) a structured inventory of urban and rural micrometeorological networks across Europe; (ii) the design and deployment of the FAIR Micrometeorological Portal, providing a digital infrastructure for data discovery, access, and standardized metadata description; and (iii) methodological guidance for quality control, gap detection, and gap filling tailored to the specific characteristics of micrometeorological time series. By providing both technical infrastructure and community-driven standards, the FAIRNESS outputs advance micrometeorological data from isolated datasets into coherent, reusable resources. Beyond technical developments, the FAIRNESS systematically addressed gaps in knowledge and skills within the micrometeorological community. A key outcome is the beginner-oriented book *Micrometeorological Measurements - An Introduction for Beginners*, which provides structured guidance on measurement design, instrumentation, data management, and quality assurance. In parallel, FAIRNESS implemented a comprehensive capacity-building programme, including summer schools, workshops, and short-term scientific missions, targeting both domain-specific competencies and transferable skills such as FAIR data stewardship, interdisciplinary collaboration, and practical problem solving. Together, these efforts contribute to strengthening the long-term usability of micrometeorological data and fostering a more integrated, FAIR-oriented research culture within the European meteorological community

# 1 Introduction

*1.1 Background and rationale*

The greatest challenges of the 21st century such as climate change, natural hazards, biodiversity and ecosystem functions, food risks, deforestation, vector born (human and animal) diseases, air quality and urbanization are either affected by or affect atmospheric conditions, particularly on the micro-scale (Shilu Tong et. All 2022, Pfenning-Butterworth et. Al 2024). Climate change is accelerating (IPCC, 2018) with the frequency and intensity of weather-related natural hazards and extreme events increasing (Alcantara-Ayala et al., 2019) requiring urgent adaptations (World Bank, 2025). This is considered both a European and a global challenge. Meteorological conditions can vary significantly within small spatial scales due to differences in surface cover, soil and topographical characteristics. A good tactical and strategic plan for facing varying weather and changing climate relies on high spatial density and the quality of site-specific observations and measurements to improve high-resolution numerical weather prediction (NWP), climate, agricultural, environmental, and urban simulation models. These challenges create the need for adequate measurement methods and strategic planning for establishing application-oriented data sets that comply with new and innovative standards.

Reliable and sufficient knowledge on environmental conditions or processes delivered from micrometeorological and microclimatological data plays a central role in assessing and modelling trends and effects of climate change (CC) and adverse weather events on the environment and ecosystems over all spatial and temporal scales. Enormous efforts have already been made on the European level in order to centralize data from ground-based (synoptic scale) and satellite measurements, weather and climate simulations and to make them available for public use (COPERNICUS, ECMWF data base, e-OBS, for example). These well-established data sources are broadly and successfully used in research, education and economics. However, beyond specific initiatives, they are still missing one very important component – micrometeorological data, i.e. data addressing meteorological conditions of microenvironment (small km scale), open and available for various application potentials and user groups.

Micrometeorological data are usually collected as a part of scientific projects and observational networks developed for different purposes, but they often "languish" in reports and institutional data storage with limited findability, interoperability, accessibility, or reusability. This presents a major lost opportunity for scientific advancement and practical applications. The current state of micrometeorological data sources reveals both considerable observational capacity and significant structural limitations. Automated Weather Stations (AWSs), operating in rural, suburban and urban areas, are widely deployed, and typically maintained by dedicated projects, specialized agencies, regional or national government offices for specific applications in the sectors of agrometeorology (e.g. pest and disease warning systems), forest, urban, and environmental meteorology (Muller et al., 2013; Lalic et al., 2020). AWSs typically measure precipitation, air temperature and relative humidity, soil temperature, water content and, depending on the specific application, also other (micro) meteorological elements such as wind speed, intensity of global and photosynthetic active radiation (PAR), and leaf wetness. The strength of the databases associated with these networks is the high spatial density of the stations representing small-scale environmental conditions and processes. The most profound weakness is a lack of findability, accessibility, interoperability and reusability of the data, i.e. they do not meet FAIR Guide Principles in managing data and metadata (Wilkinson et al., 2016). Opportunity behind existing data sets is "to define a minimal set of related but independent and separable guiding principles and practices, which enable both machines and humans to find, access, interoperate and re-use research data and metadata" (PwC EU Services, 2018). However, a possible threat is the lack of willingness of people in charge of AWS data management to join the effort and apply new procedures on existing networks, mainly due to additional costs and resource limitations.

To address these shortfalls, the FAIRNESS COST Action implemented a set of coordinated measures that resulted in the following key outcomes: a) a compiled inventory of available micrometeorological *in situ* data sets on the European level and beyond, together with gap-filling methodologies* b) the establishment of a micrometeorological knowledge share platform, hereafter referred to as the FAIR Micrometeorological Portal (FMP); c) the measurement and data management recommendations, published in the book *Micrometeorological Measurements – An Introduction for Beginners* (Lalic et al., 2026); and d) the design of a knowledge and skills enhancement plan aimed at strengthening fairness-related competencies and raising awareness of the importance of micrometeorological measurements as transferable skills† for example. The implementation and dissemination of the FMP as a data framework creates a strong background for future research and modelling studies and laid the groundwork for the development of a European micrometeorological database.

> \* The gaps which are considered refer to: missing data in time series of meteorological data and/or metadata, missing spatial coverage and lack of knowledge and expertise.
>
> † The term "fairness" will be used to describe how data meets FAIR principles (Findability, Accessibility, Interoperability and Reusability). FAIRNESS (capital letters) is the Action title.

*1.2 Objectives*

One of the major goals of the FAIRNESS Action was to develop synergy among researchers, stakeholders and civil society, to fully embrace FAIR data and "3-O" (Open data, Open innovation, Open to the World) concepts to enhance application potentials of micrometeorological data and their visibility. This was achieved through four research coordination objectives and four capacity-building objectives.

1.2.1 Research Coordination Objectives

The FAIRNESS Action was structured around research coordination objectives, aiming to improve fairness and quality of micrometeorological data across Europe and beyond where the first objective was to establish a coordinated forum of micrometeorological measurement networks and data providers. This involves identifying and connecting existing initiatives across Europe, evaluating the fairness of their datasets, and compiling an inventory of networks and data sources. By facilitating direct engagement among data holders and measurement communities, this objective supported the identification of spatial, temporal, and procedural gaps in existing datasets and practices. The resulting network of networks serves as a foundation for further collaboration, harmonization, and integration of micrometeorological data resources.

The second objective was focused on the development and implementation of the FAIR Micrometeorological Portal (FMP), which was designed both to serve as a central repository for micrometeorological data and to enable uploaded datasets to be classified as FAIR. This objective meant that the Action's micrometeorological datasets should be FAIR. The definition of FAIR (Wilkinson et al., 2016) states that data must be: Findable (F) with rich metadata and a persistent identifier, allowing both humans and machines to locate it; Accessible (A) where metadata should be retrievable using standard protocols and if data is not open, information should be provided on how to access; Interoperable (I) where data and metadata use common file formats, languages and vocabularies; and Reusable (R) where data has a usage license and clear provenance to support future reuse. Development started during its first year with workshops to discuss the functionality and specification of the system with a final agreement on system functionality and FAIR metadata description agreed at the working group meeting in Bern in May 2022. After a period of testing by members in working groups 1 and 2, the FMP was formally deployed using Amazon Web Services in October 2022.

The third objective was to build and engage a broad community of users and stakeholders through dissemination and capacity-building activities. This included the organization of summer schools, workshops, and short-term scientific missions, which served to transfer knowledge and strengthen cooperation among researchers, data managers, and policy stakeholders. The concept emphasized inclusivity by actively involving participants from less research-intensive countries and early-career researchers, thereby addressing barriers related to technical expertise and access to costly measurement technologies.

The fourth and final objective aimed to develop guidelines for beginners in micrometeorological measurements design and data management. Drawing on insights gained from the many decades of experience, community feedback, and the analysis of case studies, this objective outlined recommendations for advancing measurement techniques, promoting data standardization, and supporting future integration with FAIR-aligned infrastructures.

1.2.2 Capacity Building Objectives

In parallel with its research coordination goals, FAIRNESS invested considerable effort in capacity building to foster sustainability, knowledge exchange, and interdisciplinary integration. The first objective focused on the creation of a Pan-European, multidisciplinary network of researchers and stakeholders working in diverse fields such as meteorology and climatology, agriculture and food production, forestry, public health, urban climate, and machine learning. This objective contributed significantly to strengthening the European Research Area and fostering collaboration with experts beyond Europe, especially in emerging interdisciplinary domains such as interaction of public health and climate change.

The second objective addressed skills development and knowledge enhancement, recognizing the importance of reducing gaps that may hinder effective participation in micrometeorological research and data management. To avoid the "crowding-out" of less-resourced institutions and ensure inclusive participation, FAIRNESS designed and implemented a comprehensive training program. This included lectures, training, and other educational formats focused on data assimilation, FAIR data management, and the practical use of the FMP. The multidisciplinary character of the FAIRNESS network enabled high-quality, expert-led training sessions, while targeted engagement of early career investigators (ECIs), particularly from low-performing institutions and countries, supported their professional development and improved employability in climate and environmental sectors. An important element in the realization of this objective is development and implementation of Skill and knowledge enhancement plan.

The third capacity building objective aimed to build a FAIRNESS "neighboring" community by establishing active communication and collaboration with related projects and COST actions with overlapping interests or complementary data infrastructures.

The fourth and final objective focused on ensuring the long-term sustainability of FAIRNESS outcomes. To achieve this, an inventory of transferable activities was created to explore opportunities for incorporating selected activities, particularly those requiring limited resources, into future projects, institutional work plans, or partner-led initiatives. Additionally, the archiving of FAIRNESS outputs on ZENODO guarantees long-term data availability and visibility for at least twenty years.

## 2. Methodology

In this section, we present the methodologies that were specifically developed or adapted to align with the core objectives of the FAIRNESS Action and to support the achievement of its key outcomes. Each methodological approach reflects the interdisciplinary nature of the Action, addressing both scientific and practical challenges in micrometeorological measurements, data management, and knowledge dissemination.

*2.1 Inventory of available micrometeorological in situ data sets*

The methodology for developing the *Inventory of available micrometeorological in situ data sets* from urban and rural monitoring networks was structured to ensure efficient coordination, standardized data collection, and broad engagement across European micrometeorological stakeholders. The designed workflow consisted of the following steps:

i) **Questionnaire design**. A structured questionnaire comprising twenty-one questions is designed to gather detailed information about existing urban and rural networks and stations with multiple meteorological sensors. The questionnaire was implemented in Excel format to allow easy data entry, standardized responses, and compatibility with filtering and sorting tools for future users. It was designed to capture essential metadata on network characteristics, including station count, geographic coverage, measured variables, frequency and duration of measurements, file/data formats and contacts of responsible persons.

ii) **Stakeholder engagement and outreach**. The questionnaire was distributed to a targeted list of about 50 recipients including the International Association for Urban Climate (estimated 1000+ members) and the International Society of Biometeorology (estimated 120+ members). The outreach strategy aimed to ensure geographical diversity, cross-sector representation and both rural and urban networks inclusion for the early stages of FAIRNESS.

iii) **Data collection and pre-processing**. All responses are reviewed, checked for consistency, and prepared for integration into a representative map. A standardized template was used to harmonize submitted data, ensuring comparability across entries.

*2.2 Inventory of gap-filling methods*

The methodology adopted for gap-filling methods inventory was designed to explore, assess and recommend approaches for identifying and overcoming gaps in micrometeorological measurement series, with a particular focus on improving data assimilation and ensuring usability in research and operational applications. It included the following key steps:

i) **Selection of representative networks for gap analysis**. To ensure practical relevance and methodological diversity, two representative micrometeorological networks were selected as case studies: a rural network (PIS - Forecasting and Warning Service in Plant Protection, Serbia) and an urban network (Ghent Urban Climate Network, Belgium). These networks were chosen for their contrasting environmental settings, operational priorities, and data characteristics. Detailed information was gathered for both networks to facilitate an in-depth analysis of data gaps.

ii) **Development of a gap analysis framework**. It was designed to characterize gaps in time series database on their duration, timing, and recurrence. The analyses included identifying missing hourly data records, classifying gap durations, and evaluating temporal patterns.

iii) **Review and classification of gap-filling (GF) methods**. The methodology included review and evaluation of various gap-filling techniques, categorized into three main types based on complexity and applicability:

- Interpolation and extrapolation methods including linear regression and empirical models;
- ERA5-based methods, which involved use of debiased reanalysis data;
- Mathematical interpolation functions provided by Python including one-dimensional and multi-dimensional interpolation.
- In a separate approach, different machine learning techniques are evaluated together with traditional approaches to contrast traditional approaches with machine learning functions. A novel gap creation method was generated to provide 100% coverage in sampling the dataset while ensuring that the sampled data are randomly distributed. Gap filling across a range of different gap lengths and target variables are compared using a range of error functions. Variables selected for modelling were: mean air temperature, dew point, mean relative humidity and leaf wetness.

iv) **Testing and comparison of methods**. Identified gap-filling approaches were tested on real micrometeorological datasets with artificially created and observed data gaps. This allowed the evaluation of method performance based on accuracy metrics, with a focus on identifying suitable thresholds in terms of gap length and variable type. Also, if possible, testing was conducted over different landscapes to evaluate generalizability.

v) **Development and demonstration tools**. For easier adoption and reproducibility of gap-filling methods, tested GF methods are integrated into a user-friendly environment. The MetObs Toolkit, developed in Python for FAIRNESS 2023 Summer school, was used to demonstrate practical application of ERA5-based GF method.

vi) **Expert collaboration and documentation**. Selected GF methods were evaluated by interdisciplinary collaboration among experts in micrometeorology, data science, and environmental modelling. The resulting recommendations were the result of empirical tests, methodological comparisons, and practical experience coming from both rural and urban networks.

*2.3 FAIR Micrometeorological Portal (FMP) structure and functionality*

As one of the primary goals of the FMP is to make its datasets FAIR, it must facilitate each of the 4 F-A-I-R principles during dataset upload and dataset retrieval. When uploading micrometeorological datasets, researchers provide 3 levels of metadata: network, site and sensor metadata. This metadata supports the Accessibility, Interoperable and Reusable principles as data conforms to WMO model standards to support interoperability and the rich metadata descriptions support the reusability of data. Figure 1 shows network metadata which includes the physical location of data (Zenodo links provide persistent unique identifiers) and a number of searchable fields.

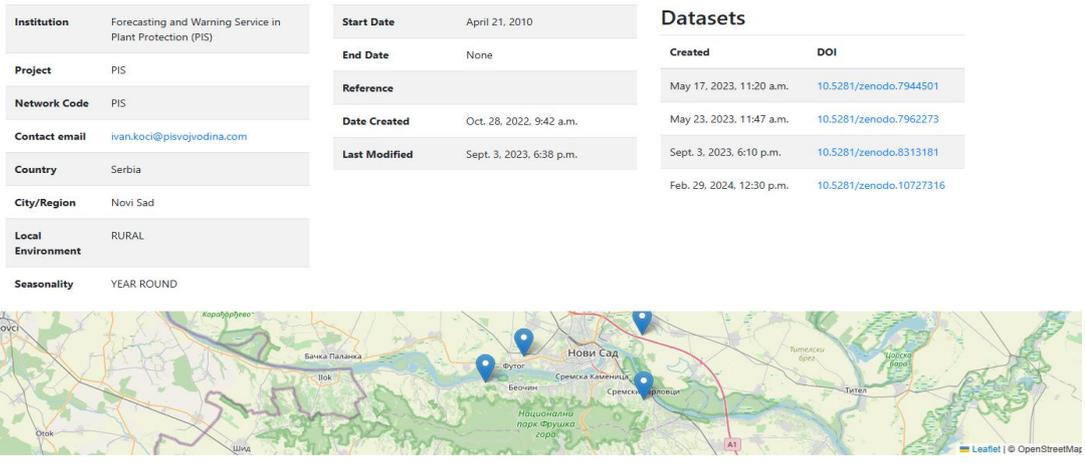

**Figure 1** Network Level Metadata

The FMP is available 24/7 with search and download facilities available to all visitors in order to deliver the Findable principle where users can search across multiple metadata levels. For example, Figure 2 displays site level metadata which enable users to decide if the dataset is suited to their analytical requirements.

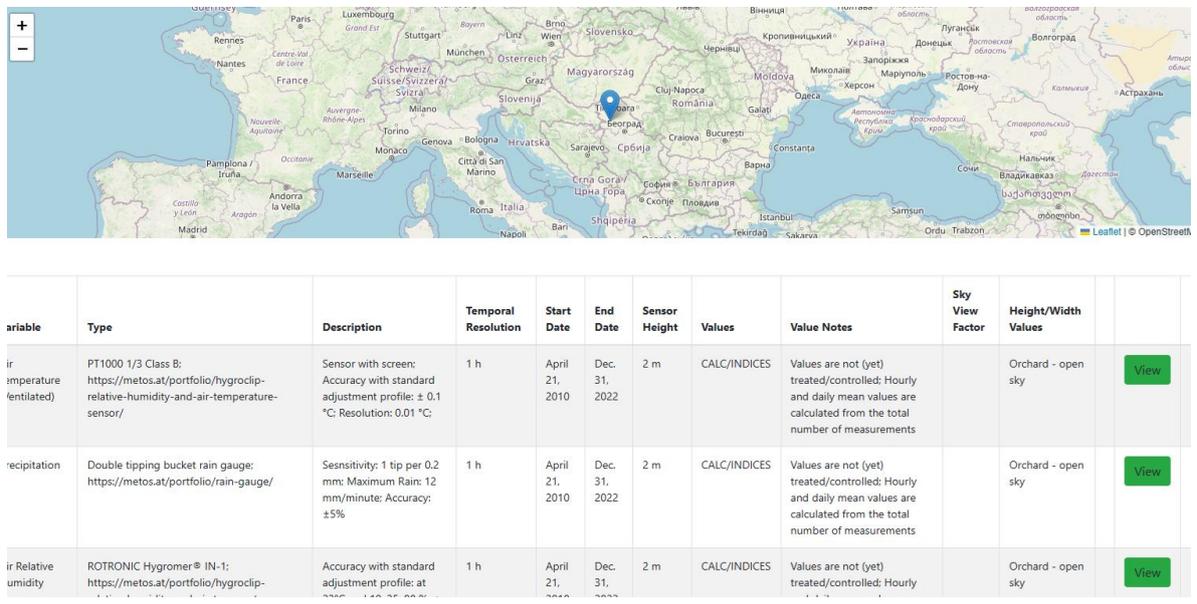

**Figure 2** Site Level Metadata

Researchers can improve the FAIRness of their data through further dissemination of data stored using the FMP. For example in Savic et. al (2023), the authors exploited the FMP's FAIR principles and Zenodo connection to provide a more extended description of the micrometeorological data.

*2.4 Knowledge and skill enhancement action plan development*

Methodology for developing *Knowledge and skill enhancement action plan* was designed to establish a structured framework for identifying, addressing and overcoming gaps in knowledge and skills relevant for micrometeorological data measurement, data management, and interdisciplinary application. Plan development included three main stages:

i) **Conceptual framing and needs assessment**. First, a critical review of emerging societal and scientific challenges requiring multidisciplinary collaboration and transferable skills was performed. This concept underscored the limitations of traditional public education models that often favor domain-specific expertise at the expense of interdisciplinary, transdisciplinary, applied and practical skill development. Within this context, both domain-specific (hard) skills and cross-cutting transferable skills were considered essential components of capacity building.

In this study, hard skills refer to domain-specific knowledge and technical competencies acquired through formal education, training, or professional experience. These skills are measurable and directly related to the ability to perform specialized tasks (e.g. instrumentations operation, data analysis). On the other hand, transferable skills are understood as competencies that can be applied across multiple roles, disciplines, or professional contexts. These may include both technical (hard) and non-technical (soft) transferable skills, depending on the level of proficiency and specific demands of a given task. A skill considered domain-specific in one profession may function as transferable in another when it supports complementary activities, interdisciplinary collaboration, or operational effectiveness. Transferability is therefore treated as context-dependent and dynamic rather than as a fixed characteristic of specific skills. The enhancement of transferable skills is considered essential for adaptability, cross-sector mobility, and long-term career development.

ii) **Structuring**. The four-element Action plan structure was adopted to guide the enhancement strategy:
   a. *assessment* - identifying individual gaps through structured self-assessment tools (questionnaire);
   b. *program design* - selecting and tailoring training methods and educational tools based on assessment results;
   c. *implementation* - realization of program activities through targeted instruments such as summer schools, short-term scientific missions, and workshops;
   d. *evaluation* - monitoring outcomes and adapting content and structure to maximize impact.

iii) **Topic definition**. Based on assessment results, core training areas are defined to structure the knowledge and skill development. These topics are used as the foundations for both hard-skill and transferable-skill development, adaptable to varied levels of participants' background and expertise.

## 3. Results

*3.1 Networks inventory and gathering of micrometeorological in situ data sets*

Through the "Inventory of available micrometeorological data and their structure" we collected data information on urban/rural micrometeorological networks across Europe, which are not part of national official networks. In total 77 urban/rural networks and individual stations across Europe are collected during the project realization, but this inventory list is not exhausted and open for extraction after the project ends. The updated inventory list, with detailed metadata and datasets information, is freely available on the following link: https://www.fairness-ca20108.eu/wg-1-outcomes/ .

The content of the inventory list ranges from individual automatic weather stations (AWSs) located on specific urban or rural locations to networks with more than 200 stations collected from 23 European countries (Fig. 3). Single AWSs are deployed like on the roof top or faculty`s yard (like in BOKU in Vienna, Austria or University of Banja Luka in Bosnia and Herzegovina, or Lisbon, etc.) or near lake or on crop lands like in Hungary, by measuring, near standard meteorological values, some additional parameters. Big networks with more than 100 stations have characteristics to cover whole urban areas like in Birmingham (UK), Zurich and Bern (Switzerland) or Rennes (France) or reach vast agricultural areas like Wegener Net in Austria, PIS in Serbia, BeRTISS network in Bulgaria, Greece and Cyprus, etc., but in most cases equipped with sensors for basic meteorological values, i.e. air temperature, soil temperature or relative humidity. During the process of creating the inventory list the relevant feedback were the networks who are currently in function, but also networks from the past (not in function anymore), and networks who are established until the end of 2023.

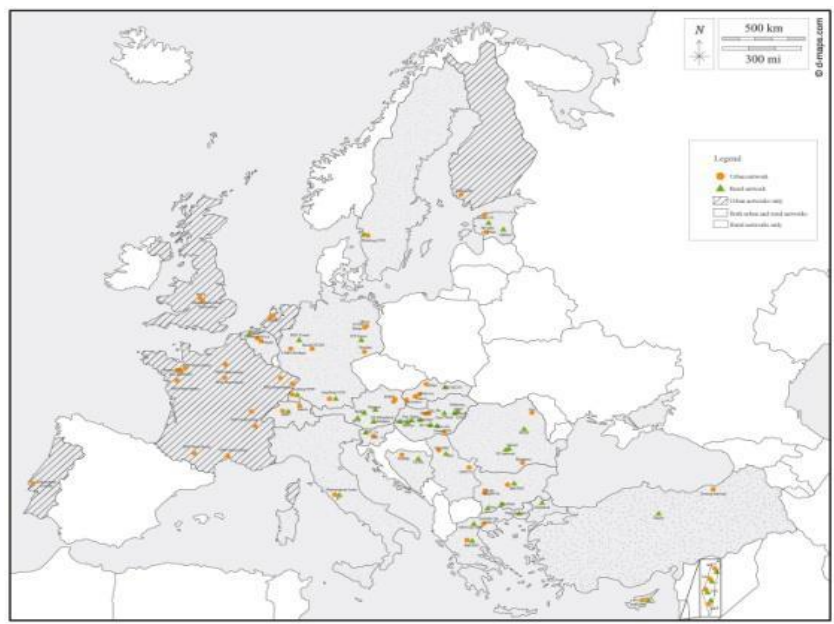

**Figure 3** Map of spatial distribution of collected urban/rural networks and stations.

Within the total number, 33 networks are declared as urban, that means all stations are located within built-up zones monitoring different meteorological values and their modifications in cities. These urban networks cover 43 cities from 16 countries. The biggest urban networks are Birmingham Urban Observatory and BUCL from Birmingham, Zurich urban heat island measurement network, Urban Climate Bern, Augsburg urban climate network from Germany, Rennes urban network from France and Parnu weather sensor network from Estonia. Rural networks are 34 in total, covering different agricultural areas from 13 countries. The rural meteorological networks with more than 100 stations are gathered from Austria (BOKU-Met and Wegener Net), BeRTISS network deployed in Bulgaria, Greece and Cyprus, ICP Forest as a pan-European forestry network and from Serbia the PIS network. Ten networks are recognized as both urban and rural. It means that part of the network is deployed in urbanized areas and part of the stations are located in the surroundings of cities that represent non-urbanized/rural micrometeorological conditions. Some of the most recognized networks are Vlinder (with more than 70 stations) that cover Brussels and rural areas across Flanders (Belgium), networks from Augsburg and Freiburg (Germany) with approximately 50 and 40 stations, respectively and a monitoring network from Goteborg (Sweden).

Working on network inventory and gathering the microclimate datasets, the intensive connections and collaborations between institutions and research groups within the FAIRNESS project have been established. This is one more important result of the FAIRNESS project obtaining a few common research papers and conference presentations. The publications are focused on station metadata explanations and making quality assessments of micrometeorological datasets from single or multiple networks, confirming that they are relevant materials for further microclimate studies. In 2023 there are two publications (Savić et al., 2023; Milošević et al., 2023) that presented the datasets from the Novi Sad Urban Network (NSUNET) and its 16 stations located in different urban areas of Novi Sad (Serbia), in addition to their potentials for urban heat island investigation, intra-urban analysis and heat adaptation assessments. By using urban temperature observations from MOCCA network in Ghent (Belgium) and TURKU network in Turku (Finland), and debiasing ERA5 reanalysis data, an innovative gap-filling algorithm was designed (Jacobs et al., 2024). This algorithm is able to reproduce the hourly and seasonal temperatures in urban environments while conserving the urban characteristics, i.e. this gap-filling method can ensure a broader utilization of urban observational datasets, without the loss of reliable detection of urban heat island patterns. Finally, (Amini et. al 2026) created the pan-European FairUrbTemp dataset that complies air temperature data from 838 street-level stations across 12 European urban networks (Amsterdam, Bern, Basel, Berlin, Biel, Birmingham, Freiburg im Breisgau, Ghent, Novi Sad, Rennes, Turku, Zurich) to test a newly developed seven steps quality control (QC) method. This QC method is specifically designed for adaptability across different networks and provides the FairUrbTemp datasets optimized for different urban studies.

During the FAIRNESS realization there were discussions at a few project meetings that near in situ measurements should be gathering the short-term monitoring datasets from mobile monitoring sensors and platforms as a further step in micrometeorological datasets standardization. Mobile monitoring is a very important method in collecting data for microclimate condition assessments, providing the measurements from specific urban or rural locations that cannot be covered in a proper way by fixed stations (Dunić et al., 2023; Lehnert et al., 2023; Arsenović et al., 2023; Vasić et al., 2024).

*3.2 Standardization of Gap Filling Approaches*

One of the goals of the Cost Action was to examine existing approaches to gap filling with the goal of creating a deterministic methodology, generating identical results from a fixed set of mathematical functions. One approach to achieving this goal is to identify a set of mathematical interpolation functions that can be applied to different gap filling requirements. A workshop in 2023 (Savić et. al. 2023b) presented approaches to both Gap Analysis methods and Gap Filling.

3.2.1 Application and Reuse of a Standard Set of Gap Filling Functions

One of the themes of the workshop was to explore if a small set of functions could provide high performance gap filling across different gap sizes and distributions and effectively become a standard methodology for filling gaps in all or most circumstances. A number of Python's interpolation functions (https://docs.scipy.org/doc/scipy/tutorial/interpolate.html) were evaluated in different test scenarios (gap sizes and with 1-dimensional and *N*-dimensional interpolation functions. Test data was prepared by uniformly selecting gap sizes (1,3,6,12,24,48 hours) with a limit of 20% missing data. The 1-D methods used (https://docs.scipy.org/doc/scipy/tutorial/interpolate.html) were the Linear, Nearest and Spline with Spline having a range of different parameter settings (Pchip, Akima and degree settings 209) in order for a suitably robust validation process. The benefits of the Nearest function was its simplicity and speed but the limitations found were poor performance with high fluctuations in missing data. The Linear function was also simple, fast and performed well when gaps were linear but underperformed with "peaky" data. The more powerful Spline function was shown to capture short-term trends better than the Linear function while also performing well with more complex curves. However, for many users it may prove difficult to understand and was found to overfit short-term trends with poor performance near the boundaries of fluctuating data. Both Akima and Pchip Interpolation functions were clearly superior to Linear and Nearest and in particular, performed far better around the boundaries in the data. However, both are quite difficult to understand without a good grounding in mathematics.

The *N*-dimensional functions facilitate the incorporation of information outside the target dataset to be used for interpolating missing data. For example, location information is used when interpolating temperature or humidity values. This study examined Python's multidimensional equivalents of the 1-D Nearest (NearestNDInterpolator) and Linear (LinearNDInterpolator) functions in addition to a more powerful Radial Basis Function (RBFInterpolator). However, for both NearestNDInterpolator LinearNDInterpolator functions, results were very similar to their 1-D equivalents. However, the RBF interpolator with its wide range of kernels offered support for more data topologies meaning that it supported many different gap distributions. On the negative side, it required high computational cost for larger datasets and for high-order kernels, can overshoot data which affects the performance of results.

3.2.2 Machine Learning Models for Gap Filling

A separate approach examined the possibility of improving the accuracy of existing gap filling methods by using machine learning functions. For a robust evaluation, this necessitated the adoption of a range of machine learning functions to deliver a comparative analysis of different approaches focusing specifically on meteorological datasets. For this study, a set of heterogeneous machine learning techniques are evaluated alongside gap filling automated the European Centre for Medium-Range Weather Forecasts atmospheric reanalyzes of the global climate, ERA-5 Land. As the machine learning models required a robust validation, a novel gap creation method was devised to provide 100% coverage in sampling the dataset while ensuring that the sampled data are randomly distributed. Gap filling across a range of different gap lengths and target variables are compared using most widely used error functions. Selected variables used for gap filling experiments included mean air temperature, dew point, mean relative humidity and leaf wetness (Lalić, B., Stapleton, A., Vergauwen, T., Caluwaerts, S., Eichelmann, E. & Roantree, M. (2024)).

Not unexpectedly, all models perform best on gap-filling temperature and dew point with worst performance on leaf wetness and model performance decreases with increasing gap length. However, the study did confirm that the machine learning functions performed best overall. In order to corroborate these findings, a detailed evaluation was provided. Micrometeorological data is multi-dimensional by nature and thus, five dimensions were used to measure and validate each of the gap filling models across 1,720 experimental configurations. The dimensions used for analysis were: 1) meteorological variables (temperature, dewpoint, humidity and leaf wetness); 2) feature sets (combinations of input variables including temporal information, data from other AWS stations, ERA-5); 3) three types of machine learning algorithms (linear regression, random forests, LightGBM) combined with two non-ML algorithms; 4) gap sizes of 1, 4, 36 and 288; and 5) geographic site locations. Each analysis starts at a one dimensional level (eg. model performance) but can become more granular as dimensions are added, eg. model performance by single target variable by gap size delivers more focused results sets, allowing for deeper interpretations around gap filling accuracy. The 4 measures used to validate predictive accuracy are R2, RMSE, nRMSE and MAE. While the discussion (Lalić, B., Stapleton, A., Vergauwen, T., Caluwaerts, S., Eichelmann, E. & Roantree, M. (2024)) focused on the higher performing models and only certain target variables, the entire set of results is available online (Roantree 2024).

A detailed discussion provided a deeper analysis of the results with one analysis examining sites, by model and by gap sizes. Using the best performing machine learning model, Light Gradient Boost (LGB) and feature set, sites were examined to determine if there was a correlation between performance and gap distribution. Here, we sought to understand if the (missing data) thresholds for sites differ in terms of gap size and a decline in the model's performance. Gap sizes of 36 and 288 hours were used for a comparative analysis. As expected, the performance of the LGB model for Temperature, in comparison to other variables, is still best with low (16%) variation in nRMSE across sites. However, significant differences were recorded for both Humidity (RH) and Dewpoint (DP) variables in both magnitude (RH=49%, DP=30%) and variability (RH=24%, DP=38%) of nRMSE. Across sites, there was an indication of poor performance (high nRMSE and RMSE) for the same locations for both RH and DP which is to be expected since DP is not a measured value but the temperature taken from (temperature) sensors at RH = 100%. This implies that DP values are affected by the operability of both thermometer and hygrometer sensors. However, as RH can be calculated using DP and TA, for locations with high nRMSE for RH, these gaps can be filled using synthesized values for both DP and TA.

Using a range of gap filling methods, the conclusion was that the selection of methods for gap filling in meteorological time series data depends not only on the inherent characteristics of the data but is also heavily influenced by the timing constraints specified by the data owner and/or user. Specifically, the urgency with which gap filling must be performed (a single day, a week, or longer), plays a crucial role in selecting the appropriate methodology. This temporal constraint influences the choice of techniques, as different methods may vary in their processing times and the immediacy with which they can address data discontinuities, thereby affecting the overall responsiveness and applicability of the gap-filling process.

*3.3 FAIR Micrometeorological Portal (FMP)*

The FMP's Search function was designed to support the Findable Accessible (*F* and *A* properties) part of FAIR. In Figure 4, the search options (country, city, local environment, seasonality, date range) are presented. Only 1 network was located as the search used quite specific parameters for matching.

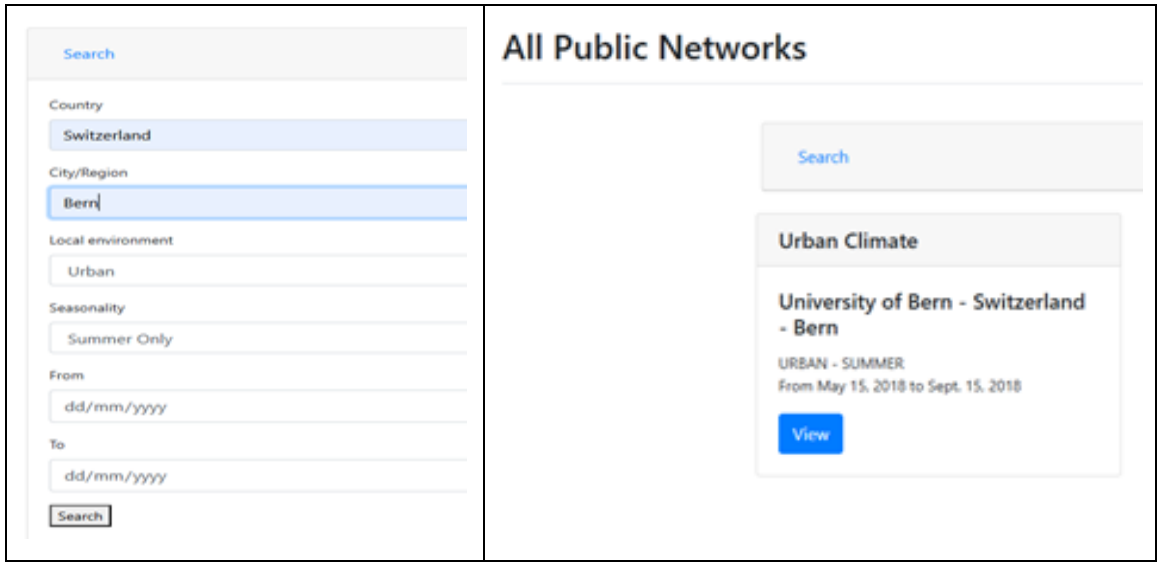

**Figure 4** FMP Search Service

3.3.1 Overview Network Distribution

Since the FMP was deployed in 2022, a total of 23 networks were created by researchers based in 16 different countries (Tab. 2), with 3 of those countries contributing more than 1 network.

| Country | TOTAL |
|---|---|
| Serbia | 5 |
| Estonia | 3 |
| Portugal | 2 |
| Austria | 1 |
| Bulgaria | 1 |
| Czech Republic | 1 |
| Finland | 1 |
| France | 1 |
| Germany | 1 |
| Greece | 1 |
| Ireland | 1 |
| Israel | 1 |
| Slovakia | 1 |
| Switzerland | 1 |
| Turkey | 1 |
| United Kingdom | 1 |

Table 2 Network Distribution by Country

In addition to a country based distribution, we can also examine distributions across Local Environment (Tab. A3) and seasonality (Tab. A4). If the Search facility from table 1 contained only 1 search parameter, Local Environment = URBAN, the results would contain 12 networks. If only Seasonality were used, then either 17 or 6 networks would be returned, according to the choice of the user.

| Local Environment | TOTAL |
|---|---|
| URBAN | 12 |
| RURAL | 7 |
| URBAN & RURAL | 4 |

Table 3 Network Distribution by Local Environment

| Seasonality | TOTAL |
|---|---|
| YEAR ROUND | 17 |
| SUMMER | 6 |

Table 4 Network Distribution by Seasonality

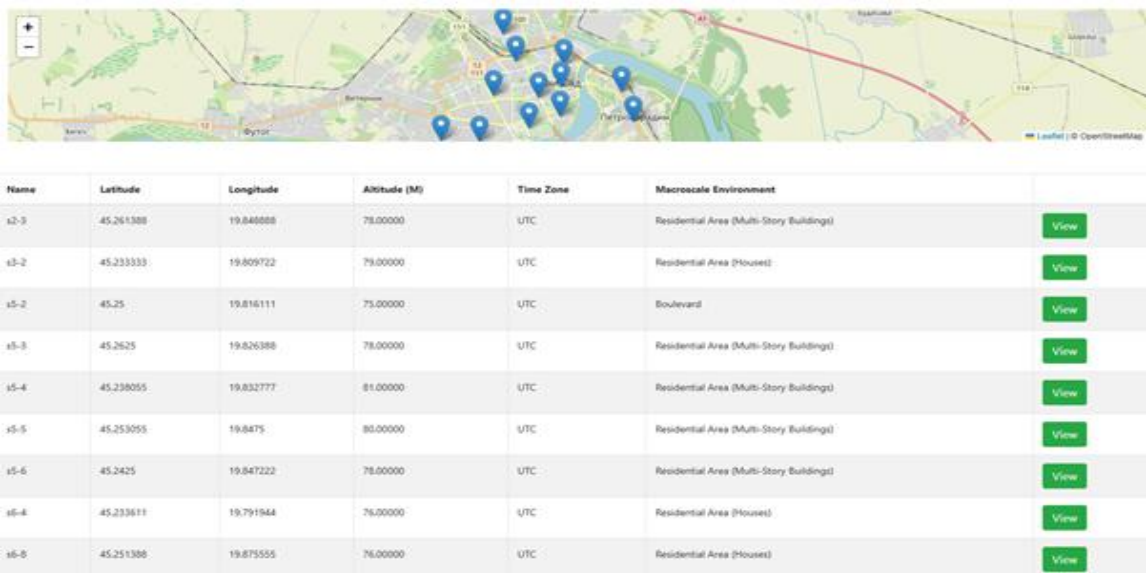

**Figure 5** Distribution of NSUNET Sites in Urban Setting

### 4.3.2 Detailed Network Information

The majority of uploaded networks have single site deployments but 3 networks are spread across multiple sites, the largest having 12 sites while the other 2 networks have 4 sites each. The largest network (Network Novi Sad Urban Network – NSUNET) is shown in Figure 5 with 12 sites contained in urban locations. The option to view sensor or setup details is available using the View option shown in Figure 5.

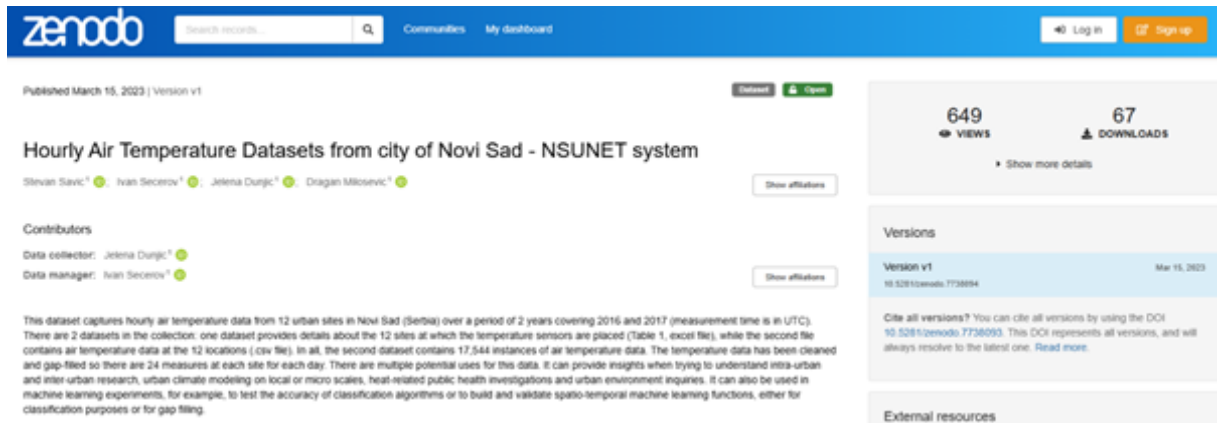

**Figure 6** NSUNET Network Data in Zenodo's Open Repository

A link field exists as part of the metadata record for this network which is present, allows the user to navigate to the actual dataset in its storage location on Zenodo. In Figure 6, information regarding the NSUNET dataset is shown as currently having 649 views and 67 downloads.

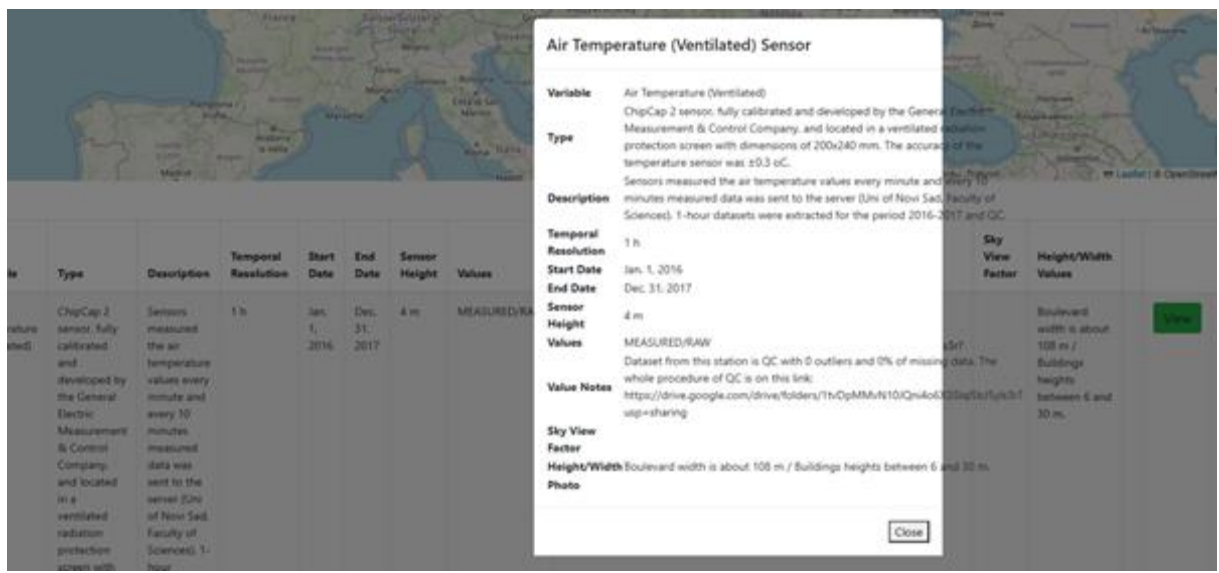

**Figure 7** Sensor Metadata and Descriptions

Sensor metadata captures details of sensor deployments for each individual site, with details to support greater reusability of (R property) of the FAIR dataset (Fig. 7). Metadata attributes were selected from the WMO standard to support higher levels of Interoperability (I property), as has been shown with unstructured data (Roantree and Liu 2014). This also enables easier integration with other datasets capturing the same properties.

*4.4. Knowledge and skill Enhancement Action Plan*

The proposed, (hard) *Knowledge and* (transferable) *skill enhancement action* plan has four stages: Assessment (A1), Program selection and design (A2), Implementation (A3), and Evaluation (A4).

4.4.1. Assessment

The first phase of the action plan (**A1**) is focused on identifying existing gaps and assessing the perceived importance of diverse competencies. This assessment utilized a specialized questionnaire to evaluate participants' awareness of hard and transferable skills. The findings based on participants' answers revealed a high level of interest among researchers in improving non-technical skills, alongside recognition that these competencies are critical for long-term career resilience in climate sciences (Fig. 8).

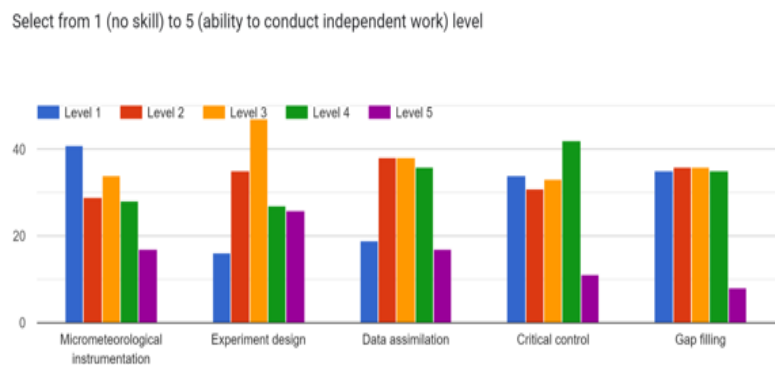

**Figure 8** Overview of micrometeorological data-related skills of participants [self-assessment]

Through this review, two distinct types of knowledge and skill gaps were identified: a) hard knowledge and technical skills gaps - particularly related to micrometeorological instrumentation, experimental design and setting and data assimilation; and b) transferrable knowledge and skill gaps - such as data interpretation, field operations, critical control, and interdisciplinary problem-solving (Fig. 9). Based on this assessment, the project identified five core technical domains (**T1–T5**) where proficiency must be integrated with broader conceptual logic:

    T1) Micrometeorological instrumentation and measurement principles;
    T2) Experimental design and field planning;
    T3) Data assimilation techniques;
    T4) Quality control;
    T5) Gap detection and filling methods.

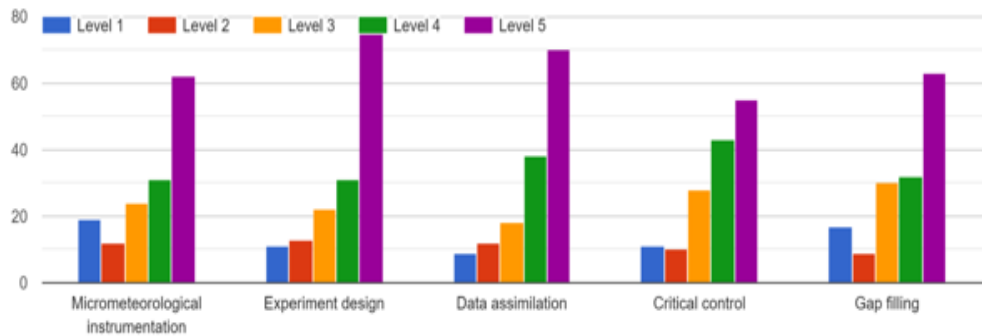

**Figure 9** Participants interest in skill enhancement

4.4.2. Program selection and design

To strengthen domain-specific knowledge and technical competencies (hard skills), a comprehensive book related to micrometeorological measurements is designed as a core component of knowledge enhancement plan. The book provides an integrated overview of theoretical foundations, experimental design principles, measurement techniques across diverse environments, data acquisition systems, and procedures for quality control and gap filing in micrometeorological data time series. Main focus is on practical guidance, standardized methods, and real-world applications in agriculture, forestry, and urban environments, making the material suitable for beginners while also serving as a reference literature for more advanced users and educators.

Transferable skills program selection and integrated design enclosed training curriculum that moves away from linear checklists toward an Integrated Transferable Skills Enhancement (TSE) approach. This design recognizes that technical proficiencies do not exist in isolation but are most effective when supported by competencies like logical synthesis and algorithmic thinking. Therefore, five core technical domains (**T1–T5**) are implemented during the FAIRNESS lifetime in the framework of:

- summer schools (https://www.fairness-ca20108.eu/training-schools/),
- workshops (https://www.fairness-ca20108.eu/workshops/) and
- short-term scientific missions (https://www.fairness-ca20108.eu/activities/short-term-scientific-missions/).

4.4.3. Implementation and Evaluation

*"Micrometeorological Measurements – An Introduction for Beginners" book overview*

The book describes in a simplified manner, supported by numerous figures and tables, summaries and highlighted notes, basic theoretical and application-oriented know-how for micrometeorological measurements (https://link.springer.com/book/10.1007/978-3-032-03884-5). It has a focus on applications in rural (agriculture, forest) and urban environments useable for beginners of specific micrometeorological measurement tasks but can serve also as a complementary information source for teachers or more sophisticated applicants.

The first chapter of the book provides a basic theoretical background on physics and processes of the atmosphere and its interactions with the earth land surface types. This starts with the short- and longwave radiation characteristics, fluxes and balance and its determining factors for absorption at the earth surface types, including a focus on seasonal and daily changes. Examples provided support a good understanding of units and realistic quantities observed under natural conditions. Based on radiation balance the transfer of energy at the absorbing surfaces is a major point for understanding environmental phenomena, described by energy partitioning in the energy balance. The next key topic for determining factors of microclimates is the understanding of transport processes of energy (heat), water (humidity) and momentum described by fluxes and its determining factors, the gradients, where the phase changes of water (e.g. evaporation, freezing, dewing) creates specific conditions and phenomena at the earth surfaces as well as in the atmosphere. The exchange of energy and water between surface (ground) and atmosphere are a key element for microclimates, so the conditions and processes of water and energy (heat) transport and storage below the surface and in the solid ground (soil) are also well introduced considering different scales in time and space.

Based on this theoretical foundation, the conditions most relevant for measurements in the atmosphere are described, including effects of conditions such as footprint, zero-place displacement, obstacles, internal boundary layers, heat sources, convection, boundary layer height and others. It is outlined that meteorological processes can be associated with typical spatial and temporal scales, which is different for other compartments of the Earth system, such as hydrological processes in soils and plant ecosystems and thus relevant for measurements settings. The question of the representativeness of meteorological measurements is especially highlighted to obtain qualitatively good and representative data. This requires that, on basis of the measuring task, the locations, the measuring principle, the measuring instrument properties and the structure and dynamics of the meteorological elements and fields must be harmonized with each other in a suitable manner. Finally, the possibilities and requirements of data transmission options and relevance are described.

The second chapter addresses in more detail the methodical recommendations for micrometeorological applications in respect to land cover types, i.e. water surfaces, bare soil, low and tall vegetation and/or classified local climate zones based on typical, often heterogenic land surface features such as in urban areas. The chapter addresses further the influence of weather conditions and surface characteristics on microclimatic parameters (e.g. footprint) and processes (e.g. convection) and the relevance for measurement setup and spatial and temporal representativeness of measurements. Similarly, most important physical soil characteristics and processes are described such as soil water and heat balance and dynamics.

In a further step the behavior of meteorological variables (such as vertical profiles) above different surface types (including within vegetated/canopy layer) as well as of variables below soil surface are described, supported by visualized examples from measurement campaigns. Here a focus is given on the important influence of vegetation on microclimate through changed surface conditions (such as roughness) and processes (such as crop transpiration). Complex surfaces, which are often met in urban areas or other complex terrain, are addressed with specific recommendations in a specific chapter, as these are often occurring application cases.

Specific methods applied for certain applications needed another special focus, including stationary, mobile, horizontal transect measurements and crowd sourcing. Related measurement standards such as typical sampling times for specific applications as well as specific challenges and recommendations, based on the authors' experiences, are outlined. Practical examples of measurements setups and results are demonstrated and visualized for a better understanding.

The final part of this chapter is dedicated to the description of up-to-date data acquisition and measurement systems, including the important issue of technical sensor characteristics, sensor calibration, maintenance of sensors, type and technics of sensor signals (analogue vs. digital) and digitalization characteristics, sensor accuracy and signal resolution and others. Specific examples and basic methods of calibration procedures are described and demonstrated. Types of data acquisition systems and datalogger designs are described and visualized in examples. Finally, types of typical ready-to-use weather station designs, available on the market, are described including their application possibilities. Final checklists accompanied by good practice examples guide the users through the complex process of planning, designing and selecting their measurement system, complete this chapter.

Chapter 3 describes good practices of measurements for single weather variables, including sensor characteristics and types following a structured scheme. It helps users quickly find the right method for them from the multitude of related sensors. With the help of notes and tables, the user can get a quick overview. First, the user is asked for the purpose/application for which the measurement of a certain variable is to be carried out (labelled with small letters), for example in subchapter for temperature "Why Should Air Temperature be Measured?" where different typical application types are listed (e.g. "Input variable for weather, climate and other models" or "Characterization of human well-being" and others). In additional tables, the user can then find out how the measuring device is to be installed for that specific application (subchapter "Where to measure") and what measurement accuracy is required for the selected application (subchapter "Which Method is to be Used for Measurement?"), where the different up-to date measurement techniques and their specifications are described, including practical recommendations. This scheme is basically followed for each relevant variable for microclimatic applications, including air and soil temperature, air humidity, wind, atmospheric pressure, short and longwave radiation, precipitation, leaf wetness and dew, soil temperature and soil heat flux, soil moisture and soil water tension. In further, there are summarizing sections describing in the same structure combination sensors, evapotranspiration measurements and calculations, flux measurements and chamber methods. Where suitable, links to the other chapter contents (such as to Chapter 2 on the described general methods) or further literature sources are provided or recommended for further in-depth reading.

The fourth chapter is finally dedicated to "Quality control and recovery of meteorological data" which is a crucial activity of postprocessing received raw data from meteorological sensors or stations for further analysis or ensuring application potentials. Meteorological measurements require significant time, human effort, and resources, including instrumental and operational costs. The resulting meteorological database must have a reliable consistency, which must be designed according to the user's needs. The whole measurement process, from planning and operational phases to the development of the complete dataset, must be accompanied by continuous quality control (QC) and quality assurance (QA) activities. Related practices are vital to correctly identifying, addressing, and correcting errors that may occur during data collection, processing, and analysis. The integrity and representativeness of meteorological datasets need to be maintained, despite challenges posed by instrumentation limitations, environmental conditions, and human errors. In that context, a main important aspect in this chapter is the description of the importance and categories of meta-data, including a practical checklist of recommendations.

The subchapters provide a comprehensive overview on the type (e.g. gaps, inconsistencies, unreliable data, time shifts and many others) and sources of data errors (e.g. installation, observer, transmission errors and others) as well as quality control methods with practical examples. Necessary steps and best practices for the QA/QC of micrometeorological data are described accordingly. Finally, methods of data correction, homogenization, and gap-filling are presented, including practical examples. By describing standardized methodologies and alternative data sources where appropriate, this chapter provides a framework for enhancing the reliability and applicability of micrometeorological data across a wide range of applications.

*4.5 Micrometeorological Measurements as Transferable Skills*

The evaluation results from the FAIRNESS Summer Schools conducted between 2022 and 2025 demonstrated consistently high participant satisfaction and confirm the strong scientific and educational impact of the training activities. In line with the FAIRNESS commitment to continuous improvement, we performed a systematic analysis of the participant evaluation reports to refine the strategic structure of our training activities. The FIRST Summer School in 2022 was designed as a comprehensive program covering all primary topics (T1–T5) for both agricultural and urban micrometeorology; however, the feedback indicated that this combined approach proved too broad to allow for the necessary focus on practical topics and detailed scientific exploration. Based on the 2022 evaluation findings, we implemented significant structural changes, splitting the curriculum into specialized editions: the urban-focused school in Ghent (2023), the urban agrometeorology-focused sessions in Budapest (2024) and agrometeorology focused in Ancona (2025), while substantially increasing the ratio of hands-on exercises to theoretical lectures. This adaptive management ensured that subsequent schools provided the specialized technical mastery and intensive practical training requested by the research community.

The implementation of training program followed a 30-20-50 distribution in respect to lectures – demonstration - hands-on training. The 172 participants attending the four SSs served as the primary boost for skill integration and enhancement for PhD students, early career scientists, and selected experts. Overcoming technical gaps was achieved through **19** STSMs for individual intensive training and **6** specialized WSs with 120 participants focusing on instrumentation, FAIR data, protocols and gap filling techniques.

The final stage, **Evaluation**, measured the efficacy of this integrated approach. Detailed results of evaluation could be found on the FAIRNESS web page (https://www.fairness-ca20108.eu/training-schools/). A post-training questionnaire administered to training participants demonstrated a significant increase in confidence across the technical spectrum (Fig. 10).

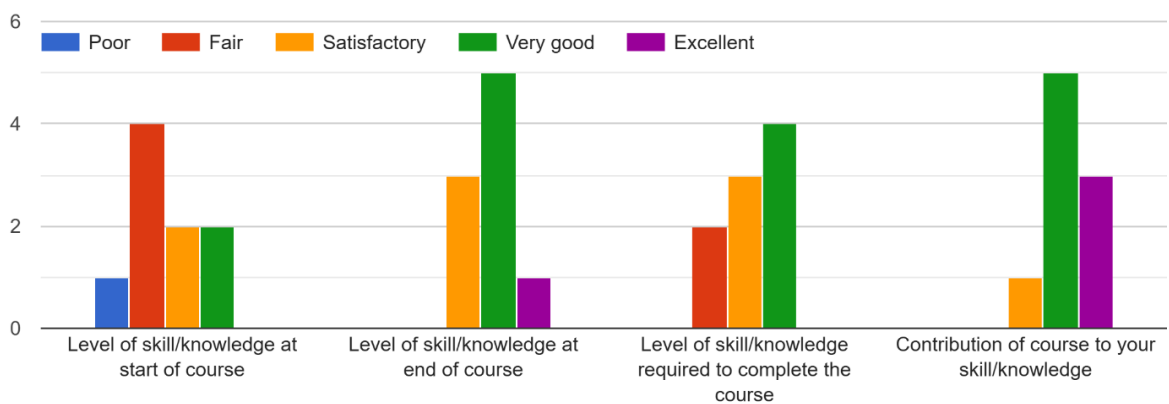

**Figure 10** Contribution to learning: Participants self-assessment of proficiency (Pre- vs. Post-Summer School) for T1–T5 categories

## 4. Discussion

The FAIRNESS generated substantial scientific, technological, and socioeconomic impacts, both in the short and long term.

From a scientific perspective, FAIRNESS has contributed to improved know-how regarding *in situ* micrometeorological data, particularly for weather- and climate-related impact studies and the calibration and validation of existing models. The availability of high-resolution, site-specific data can enhance the verification of numerical weather prediction models on small spatial scales, especially for extreme and adverse weather events. In the long term, FAIRNESS lays the foundation for establishing a European micrometeorological database, which will support more comprehensive and integrated research across scales and disciplines. The FMP 2.0, explicitly focused on assessing and improving the FAIRness of micrometeorological data, will play a key role in enabling better data findability, accessibility, interoperability and reusability, supporting applications in agriculture, forestry, urban climate studies, urban planning, and climate services.

The technological impact of FAIRNESS includes the introduction and refinement of standardized methods for gap filling in meteorological datasets (Lalic et al., 2024, Jacobs et al., 2024, Vergauwen et al., 2024) and the launch of FMP 2.0 (Roantree et al., 2023). The FMP 2.0 is now strategically dedicated to enhancing the FAIR character of datasets, helping data owners align with best practices and facilitating future integrations with open data infrastructures. In the long term, FAIRNESS is expected to contribute to the Open Research Europe (https://open-research-europe.ec.europa.eu/for-authors/data-guidelines/) and foster technological innovation by enabling improved model calibration, more robust early warning systems, and more efficient urban energy and public health strategies, all powered by high-quality, reusable data.

In terms of socioeconomic impact, FAIRNESS has strengthened transdisciplinary and transnational networking, increased the visibility of underrepresented institutions and countries, and raised awareness among the public and policymakers about the real-world impacts of weather and climate (Firanj Sremac & Lalic, 2023, Viet Cuong, 2025). To maximize its impact, FAIRNESS has implemented several strategic measures across three dimensions: knowledge creation, knowledge transfer, and career development.

In terms of knowledge creation, FMP 2.0 has the potential to contribute to the EU data economy and open data market while improvement and harmonization of existing measurement guidelines will help to enhance both the quality and quantity of micrometeorological data in future applications. Open-access publication of FAIRNESS outcomes ensures broad dissemination and uptake of the generated knowledge (Nikolova and Matev 2023) (Lalic et al. 2026).

In terms of career development, FAIRNESS actively supported the acquisition of transferrable and interdisciplinary skills, which are increasingly valued in both academia and the labor market. Through its training activities and collaborative writing, the Action enhances competencies in creative problem solving, interdisciplinary integration, and practical aspects of measurement design and implementation. For early career investigators, participation in FAIRNESS was often their first exposure to working within international and multidisciplinary teams. For experienced researchers, the Action provided opportunities to expand their collaborative network, FAIR data stewardship and the application of micrometeorological observations, thus improving their capacity to further exploit and extend the results of FAIRNESS.

## 5. Conclusions

For the first time, a comprehensive, structured inventory of urban and rural networks/stations that monitor local and micro-scale climate conditions, FAIR micrometeorological knowledge share platform, FMP 2.0 of site-specific measurements, and beginners guide for micrometeorological measurements are established on the European level addressing the primary needs of: i) research applications (e.g. risk assessment and management decision-supporting tools under climate change conditions while considering the greatly varying ecosystem conditions in Europe); ii) stakeholders (specialists/experts, manufacturers of equipment, pharmaceuticals and food industry), policy and decision makers (from local to national and international level); and iii) specialized user groups and general public related/interested in meteorology/climatology, agriculture and forestry, environmental sciences and health issues.

The scientific, technological and societal measures implemented within FAIRNESS demonstrate the Action's strong potential to generate meaningful and sustained innovation. By building interdisciplinary networks, developing new tools, improving data management standards, and fostering capacity across generations of researchers, FAIRNESS has established a strong foundation for long-term improvements in the use of micrometeorological data for climate adaptation, environmental protection, and evidence-based policymaking. However, as is the case with many capacity-building, research-coordination initiatives, the full impact of these efforts is not always immediate. Outcomes such as enhanced data sharing, standardized methodologies, and broader uptake of FAIR principles often emerge gradually, as tools and knowledge are integrated into ongoing research projects, institutional practices and education. This is especially the case for early-career researchers whose involvement in such Actions will shape their professional development, research focus and future leadership roles. As a result, the legacy of FAIRNESS is expected to extend well beyond the Action's formal duration, through its contributions to a more open, collaborative, and FAIR-oriented research culture in Europe and beyond.

**Acknowledgements**

Mark Roantree and Michael Scriney are supported by Taighde Éireann – Research Ireland under Grant number 12/RC/2289_P2.